\documentclass{aastex62}
\usepackage{amsmath,amssymb,CJK}

\received{--}
\revised{--}
\accepted{--}
\submitjournal{AJ}

\shorttitle{Pre-discovery Activity of 2I/Borisov Beyond 5~AU}
\shortauthors{Ye et al.}

\begin{document}
\begin{CJK*}{UTF8}{gbsn}

\title{Pre-discovery Activity of New Interstellar Comet 2I/Borisov Beyond 5~AU}

\correspondingauthor{Quanzhi Ye}
\email{qye@umd.edu}

\author[0000-0002-4838-7676]{Quanzhi Ye (叶泉志)}
\affiliation{Department of Astronomy, University of Maryland, College Park, MD 20742}

\author[0000-0002-6702-7676]{Michael S. P. Kelley}
\affiliation{Department of Astronomy, University of Maryland, College Park, MD 20742}

\author{Bryce T. Bolin}
\affiliation{IPAC, California Institute of Technology, 1200 E. California Blvd, Pasadena, CA 91125}

\author[0000-0002-2668-7248]{Dennis Bodewits}
\affiliation{Physics Department, Leach Science Center, Auburn University, Auburn, AL 36849}

\author[0000-0003-0774-884X]{Davide Farnocchia}
\affiliation{Jet Propulsion Laboratory, California Institute of Technology, 4800 Oak Grove Drive, Pasadena, CA 91109}

\author{Frank J. Masci}
\affiliation{IPAC, California Institute of Technology, 1200 E. California Blvd, Pasadena, CA 91125}

\author[0000-0002-2058-5670]{Karen J. Meech}
\affiliation{Institute for Astronomy, University of Hawaii, 2680 Woodlawn Drive, Honolulu, HI 96822}

\author{Marco Micheli}
\affiliation{ESA NEO Coordination Centre, Largo Galileo Galilei, 1, 00044 Frascati (RM), Italy
}
\affiliation{INAF - Osservatorio Astronomico di Roma, Via Frascati, 33, 00040 Monte Porzio Catone (RM), Italy}

\author[0000-0002-0439-9341]{Robert Weryk}
\affiliation{Institute for Astronomy, University of Hawaii, 2680 Woodlawn Drive, Honolulu, HI 96822}

\author[0000-0001-8018-5348]{Eric C. Bellm}
\affiliation{DIRAC Institute, Department of Astronomy, University of Washington, 3910 15th Avenue NE, Seattle, WA 98195}

\author{Eric Christensen}
\affiliation{University of Arizona, Lunar and Planetary Laboratory, 1629 E. University Blvd., Tucson, AZ 85721}

\author{Richard Dekany}
\affiliation{Caltech Optical Observatories, California Institute of Technology, Pasadena, CA 91125}

\author{Alexandre Delacroix}
\affiliation{Caltech Optical Observatories, California Institute of Technology, Pasadena, CA 91125}

\author[0000-0002-3168-0139]{Matthew J. Graham}
\affiliation{Division of Physics, Mathematics, and Astronomy, California Institute of Technology, Pasadena, CA 91125}

\author[0000-0001-5390-8563]{Shrinivas R. Kulkarni}
\affiliation{Division of Physics, Mathematics, and Astronomy, California Institute of Technology, Pasadena, CA 91125}

\author[0000-0003-2451-5482]{Russ R. Laher}
\affiliation{IPAC, California Institute of Technology, 1200 E. California Blvd, Pasadena, CA 91125}

\author[0000-0001-7648-4142]{Ben Rusholme}
\affiliation{IPAC, California Institute of Technology, 1200 E. California Blvd, Pasadena, CA 91125}

\author{Roger M. Smith}
\affiliation{Caltech Optical Observatories, California Institute of Technology, Pasadena, CA 91125}



\begin{abstract}
Comet 2I/Borisov, the first unambiguous interstellar comet ever found, was discovered in August 2019 at $\sim3$~au from the Sun on its inbound leg. No pre-discovery detection beyond 3~au has yet been reported, mostly due to the comet's proximity to the Sun as seen from the Earth. Here we present a search for pre-discovery detections of comet Borisov using images taken by the Catalina Sky Survey (CSS), Pan-STARRS and Zwicky Transient Facility (ZTF), with a further comprehensive follow-up campaign being presented in \citet{Bolin2019}. We identified comet Borisov in ZTF images taken in May 2019 and use these data to update its orbit. This allowed us to identify the comet in images acquired as far back as December 2018, when it was 7.8~au from the Sun. The comet was not detected in November 2018 when it was 8.6~au from the Sun, possibly implying an onset of activity around this time. This suggests that the activity of the comet is either driven by a more volatile species other than H$_2$O, such as CO or CO$_2$, or by exothermic crystallization of amorphous ice. We derive the radius of the nucleus to be $<7$~km using the non-detection in November 2018, and estimate an area of $\sim0.5$---$10~\mathrm{km^2}$ has been active between December 2018 and September 2019, though this number is model-dependent and is highly uncertain. The behavior of comet Borisov during its inbound leg is observationally consistent with dynamically new comets observed in our solar system, suggesting some similarities between the two.
\end{abstract}

\keywords{comets: individual (2I/Borisov)}


\section{Introduction}

Comet 2I/Borisov\footnote{Formerly designated as C/2019 Q4 (Borisov). See \href{https://www.minorplanetcenter.net/mpec/K19/K19RA6.html}{Minor Planet Electronic Circular (MPEC) 2019-R106} and \href{https://www.minorplanetcenter.net/mpec/K19/K19S72.html}{MPEC 2019-S72}.} (2I for short), discovered by G. Borisov at the Crimean Astrophysical Observatory on 2019 August 30, is the second interstellar object and the first unambiguous interstellar comet ever found. Early observations of 2I have revealed a sizeable coma \citep{deLeon2019,Guzik2019,Jewitt2019b} and the detection of the emission from CN and atomic oxygen \citep{Fitzsimmons2019,Opitom2019,McKay2019}. This makes it distinctively different from the first discovered interstellar object 1I/`Oumuamua, which did not exhibit a detectable cometary feature such as a coma and/or a tail \citep{Meech2017,Ye2017,ISSI2019}. The detection of activity is important as it provides a way to probe the composition of the nucleus, which can be done through a direct analysis of gases in the coma, and by observing how the activity responds to different levels of insolation at different heliocentric distances.

2I was discovered at a heliocentric distance $r_\mathrm{H}=2.98$~au, with a brightness ($V=18$~mag) that is well within the reach of modern near-Earth object (NEO) surveys. The belated discovery is primarily due to the fact that the comet was within $45^\circ$ from the Sun, within the typical solar avoidance zone for ground-based optical telescopes, from early May to early September 2019.

A detection (or non-detection) at $r_\mathrm{H}\gg3$~au would better constrain the size of the nucleus (as the comet is likely less active) as well as its inbound orbit, and would be diagnostic of the comet's volatile composition \citep{Fitzsimmons2019}, since the activity of most known (i.e., solar system) comets within $\sim3$--5~au is driven by water ice sublimation \citep[e.g.,][]{Meech2004}. All-sky, time-domain surveys allow for serendipitous observations of objects before they are discovered. Here we present a search for pre-discovery detections of 2I in the data of several sky surveys. We will describe the search process in \S~2, the procedure of photometric analysis in \S~3, and will then discuss the implication of the detections and non-detections in \S~4 and \S~5.

\section{Pre-discovery Detections}
\subsection{Zwicky Transient Facility}
\label{sec:predisc:ztf}

The Zwicky Transient Facility (ZTF) is a wide-field optical survey utilizing the 1.2 m Palomar Oschin Schmidt and a dedicated camera with 55 deg$^2$ field-of-view \citep{Bellm2019,Graham2019,Masci2019}. Apart from two baseline surveys and a number of mini-surveys, ZTF executes a mini-survey (``Twilight Survey'') that observes regions with solar elongation down to $35^\circ$ \citep{Ye2019}, which is particularly suitable for the search of pre-discovery detections of 2I. The first phase of the ZTF Twilight Survey was executed from November 2018 to June 2019, fortuitously covering the period before 2I reached solar conjunction in July 2019.

By using the {\tt ZChecker} comet monitoring package \citep{Kelley2019} and JPL orbit solution \#12 (the most recent JPL solution at the time that the search was conducted), we identified a total of 202 images from 2018 October 1 to the discovery date of 2I (2019 August 30) that potentially contained the comet. All images use 30~s exposure times, and were taken in a wide range of observing circumstances, with $5\sigma$ limiting magnitudes between 16--21 and seeing between $1''$--$5''$. The comet reached solar conjunction in July of 2019, and the last three sets of images acquired before that were taken in the course of the Twilight Survey on 2019 April 29, May 2 and May 5 respectively, when 2I was at $r_\mathrm{H}=5.20$--5.09~au. Prior to these Twilight Survey data, the last image was taken on 2019 April 16, when the comet was at $r_\mathrm{H}=5.45$~au. The April 29 and May 5 observations were not optimal because only a few images were acquired, high sky background, and/or the comet being near image edges. The May 2 observations were particularly fortuitous as the comet was located in an overlapping strip between two fields, and thus have $2\times$ more images than we typically have (8 images total, instead of the typical 4 images per field). Therefore, we focused on the 2019 May 2 data for our initial search of a pre-discovery detection.

\begin{figure*}
\includegraphics[width=\textwidth]{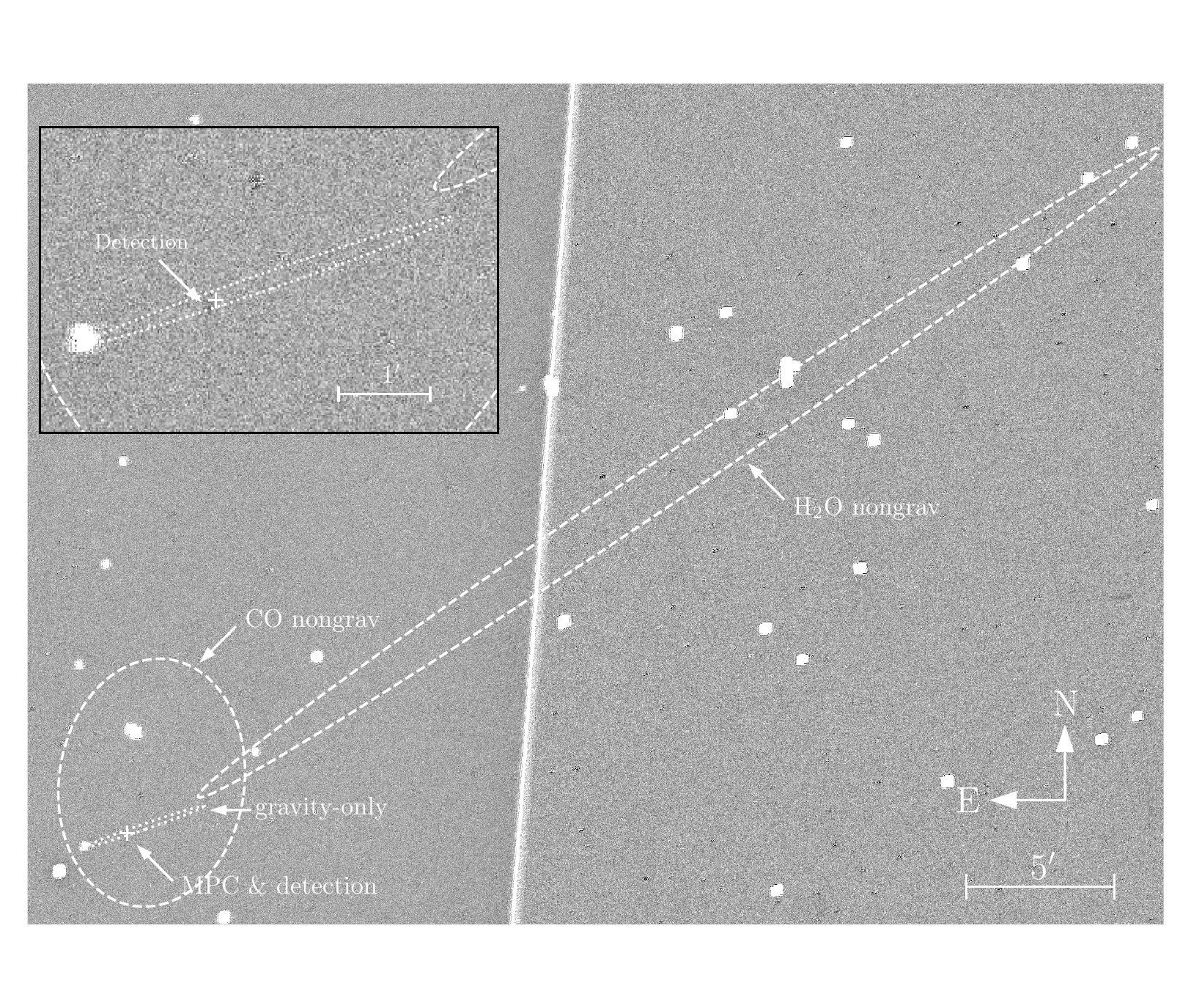}
\caption{Combined image stacked following the predicted motion of 2I, using 8 images taken on 2019 May 2, with the pre-discovery detection of the comet marked. The MPC position is calculated using the orbit published in \href{https://www.minorplanetcenter.net/mpec/K19/K19T44.html}{MPEC 2019-T44}. The motion rate difference between different orbit solutions is smaller than 1/5th of a pixel over the entire imaging session and is negligible. The input images have been subtracted with the reference images to remove background stars \citep[i.e. ``differenced'' images, see][]{Masci2019}. The white blobs are masked stars. The image is plotted in inverted linear scale (i.e. sky is white, the comet is dark.}
\label{fig:stack}
\end{figure*}

The astrometric observations available up to this point did not provide meaningful constraints on the cometary non-gravitational perturbations \citep[e.g.,][]{Yeomans2004}, which lead to different orbit solutions. To capture the ephemeris variations using different model assumptions, we fit the available astrometry as of 2019 October 3 using a gravity-only model, and two non-gravitational models that either assumed sublimation of H$_2$O \citep{Marsden1973} or CO (with sublimation rate $\propto r_\mathrm{H}^{-2}$)\footnote{To our best knowledge, the only CO-driven non-gravitational model published to-date is the \citet{Yabushita1996} model. However, their elbow at $\sim5$~au is inconsistent with both theoretical and observational results, which suggests an elbow of $\gg10$~au \citep[e.g.][Hui et al. in prep]{Biver1996, Gunnarsson2002, Meech2004}. Here we simply assume the sublimation rate follows $\propto r_\mathrm{H}^{-2}$, since the elbow is likely much larger than the largest $r_\mathrm{H}$ in our dataset.} to be the primary driver of comet activity. We then combined all 8 images from May 2 following the apparent motion of 2I. The three solutions are tabulated in Table~\ref{tbl:solutions} (together with the final orbit, after we have successfully identified the comet in the ZTF pre-discovery data, as we discuss below), and the combined image as well as the uncertainty ellipses of the three solutions are shown as Figure~\ref{fig:stack}. Because of the short arc and the low elongation of many of the astrometric positions, we note that the estimated non-gravitational parameters could be unreliable. In particular, the parameters for the H$_2$O-driven model appear to be too large to be credible, and are possibly caused by astrometric biases at such a small solar elongation.

\begin{table*}
\begin{center}
\caption{Orbital elements of various solutions in Figure~\ref{fig:stack}, as well as the best solution (JPL 37), which uses precovery data and a CO-driven non-gravitational forces. \label{tbl:solutions}}
\begin{tabular}{lcccc}
\hline
Parameter & Gravity-only & Nongrav -- H$_2$O & Nongrav -- CO & Final -- CO \\
 & (without precovery) & (without precovery) & (without precovery) & (with precovery) \\
\hline
Epoch (TDB) & 2019 Sep 16.0 & 2019 Sep 16.0 & 2019 Sep 16.0 & 2019 Dec 20.0 \\
Perihelion time (TDB) & 2019 Dec 8.65 $\pm$ 0.07 d & 2019 Dec 13 $\pm$ 1 d& 2019 Dec 9.1 $\pm$ 0.4 d & 2019 Dec 8.551 $\pm$ 0.001 d\\
Perihelion distance $q$ (au) & $2.003\pm0.003$ & $1.85\pm0.05$ & $1.98\pm0.02$ & $2.00664\pm0.00004$ \\
Eccentricity $e$ & $3.34\pm0.01$ & $2.7\pm0.2$ & $3.27\pm0.07$ & $3.3576\pm0.0003$ \\
Inclination $i$ & $44.08^\circ\pm0.03^\circ$ & $45.6^\circ\pm0.5^\circ$ & $44.3^\circ\pm0.2^\circ$ & $44.0515^\circ\pm0.0004$ \\
Long. Ascending Node $\Omega$ & $308.12^\circ\pm0.03^\circ$ & $306.7^\circ\pm0.4^\circ$ & $308.0^\circ\pm0.2^\circ$ & $308.1488^\circ\pm0.0004^\circ$ \\
Argument of Perihelion $\omega$ & $209.21^\circ\pm0.06^\circ$ & $213^\circ\pm1^\circ$ & $209.7^\circ\pm0.5^\circ$ & $209.1227^\circ\pm0.0009^\circ$ \\
Radial accel. $A_1$ ($\mathrm{au~day^{-2}}$) & - & $(3.5\pm0.7)\times10^{-4}$ & $(1.4\pm1.5)\times10^{-5}$ & $(-4.4 \pm 3.2)\times 10^{-8}$ \\
Transverse accel. $A_2$ ($\mathrm{au~day^{-2}}$) & - & $(-2.7\pm0.3)\times10^{-4}$ & $(-6.2\pm5.6)\times10^{-6}$ & $(1.3 \pm 0.3)\times 10^{-7}$ \\
\hline
\end{tabular}
\end{center}
\end{table*}

\begin{figure*}
\centering \includegraphics[width=0.8\textwidth]{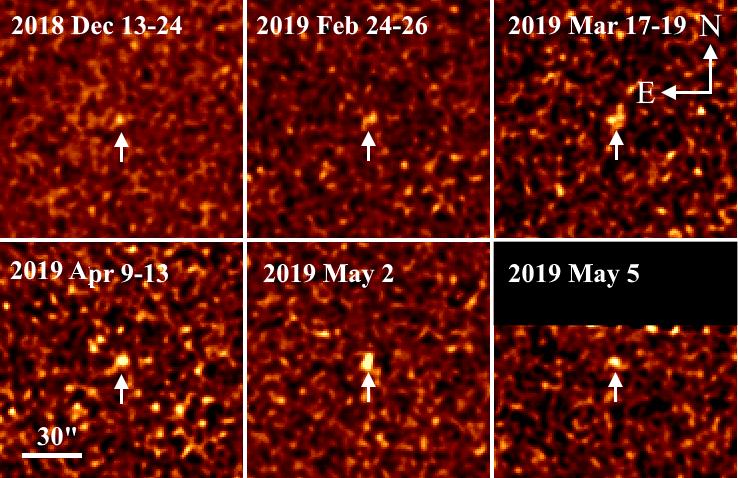}
\caption{A close-up view of the pre-discovery detection of 2I in ZTF data. To improve the clarity, the stamps have been smoothed by a moving Gaussian function with a width of 7~pix at $1.5\sigma$. The stamps are plotted in hyperbolic sine scale.}
\label{fig:stack-stamp}
\end{figure*}

We identified a possible detection at a signal-to-noise ratio ($S/N$) of $\sim10$ on the combined May 2 image (Figure~\ref{fig:stack-stamp}). The detection is within the $3\sigma$ uncertainty ellipses of gravity-only and CO non-gravitational solutions, and about $5''$ southeast of the predicted Minor Planet Center (MPC) position. Visually, the object is about $\sim5''$ in diameter and is largely circular in shape, with no apparent sign of a tail. The object is barely visible on individual frames (Figure~\ref{fig:stack-stamp}), with a motion consistent with that of 2I.

We then examined other images in the time period in question, using the same shift-and-stack technique as outlined above. The detection on 2019 May 2, if real, would have greatly reduced the ephemeris uncertainty from a few arcminutes to a few arcseconds back to January 2019, and would enable more pre-discovery data to be found. Since the comet is extremely faint, in order to eliminate contamination due to variable sky conditions and passing background stars, we only use frames that have (1) $5\sigma$ limit of $r\gtrsim19.5$, (2) average full-width-half-maximum (FWHM) of $<3$ pixels, and (3) have no background stars that are within $10''$ from the predicted position of the comet.

All the pre-discovery detections and non-detections are summarized in Table~\ref{tbl:prediscovery}. We were able to trace the object back to 2018 December 13. Apart from the 2019 May 2 data, the object is not visible in individual frames, and a clear detection usually requires stacking images from multiple nights. By including these astrometric positions\footnote{Published in \href{https://www.minorplanetcenter.net/mpec/K19/K19V34.html}{MPEC 2019-V34} and \href{https://www.minorplanetcenter.net/mpec/K19/K19W50.html}{MPEC 2019-W50}.} with the post-discovery astrometric measurements of 2I and considering the non-gravitational effect, we were able to get a satisfactory orbital fit with residuals of order $1''$, which is slightly above the average compared to typical ZTF astrometry (better than $0.5''$). This is due to a weak systematic bias in the data, which is possibly caused by the fact that most astrometric data were taken at low solar elongation and therefore at high airmass, introducing some differential color refraction (DCR) bias (see also the discussion in \S~\ref{sec:disc:nongrav}). Nevertheless, we identify the object observed from 2018 December 13 to 2019 May 5 as 2I.

\begin{table*}
\begin{center}
\caption{Summary of all the pre-discovery observations. The heliocentric and the geocentric distances of the comet ($r_\mathrm{H}$ and $\varDelta$) are given at the median time of the image epochs. FWHM is the average full-width-half-maximum of the image. \label{tbl:prediscovery}}
\begin{tabular}{ccccccccc}
\hline
Images date & Median date (UT) & Survey & $r_\mathrm{H}$ (au) & $\varDelta$ (au) & Images used & FWHM & Res. & $r_\mathrm{PS1}$ mag \\
\hline
2018 Oct 31 -- 2018 Nov 21 & 2018 Nov 8.82 & ZTF & 8.55 & 7.90 & $g$: 12; $r$: 16 & $1.4''$--$3.0''$ & - & $m_{3\sigma}>22.69$ \\
2018 Dec 13 -- 2018 Dec 22 & 2018 Dec 19.15 & ZTF & 7.75 & 6.99 & $g$: 6; $r$: 6 & $1.9''$--$2.7''$ & $0.7''$ & $21.19\pm0.15$ \\
2019 Jan 17 & 2019 Jan 17.30 & PS1 & 7.18 & 6.58 & $i$: 4 & $0.9''$--$1.1''$ & - & - \\
2019 Feb 24 -- 2019 Feb 26 & 2019 Feb 25.18 & ZTF & 6.42 & 6.26 & $g$: 3; $r$: 4 & $1.6''$--$3.6''$ & $0.6''$ & $20.97\pm0.18$ \\
2019 Mar 1 & 2019 Mar. 1.10 & CSS & 6.35 & 6.24 & Clear: 4 & $\sim3''$ & $1.1''$ & $\gtrsim21$ \\
2019 Mar 16 -- 2019 Mar 18 & 2019 Mar 17.18 & ZTF & 6.02 & 6.14 & $g$: 4; $r$: 5 & $2.1''$--$2.9''$ & $1.5''$ & $20.46\pm0.16$ \\
2019 Apr 9 -- 2019 Apr 13 & 2019 Apr 12.16 & ZTF & 5.53 & 5.97 & $g$: 3; $r$: 3 & $1.6''$--$3.0''$ & $1.4''$ & $20.42\pm0.09$ \\
2019 May 2 & 2019 May 2.16 & ZTF & 5.15 & 5.79 & $r$: 8 & $2.0''$--$2.1''$ & $1.1''$ & $20.12\pm0.11$ \\
2019 May 5 & 2019 May 5.15 & ZTF & 5.09 & 5.76 & $r$: 4 & $1.8''$--$2.5''$ & $0.7''$ & $20.16\pm0.21$ \\
\hline
\end{tabular}
\end{center}
\end{table*}

\begin{figure*}
\centering \includegraphics[width=0.8\textwidth]{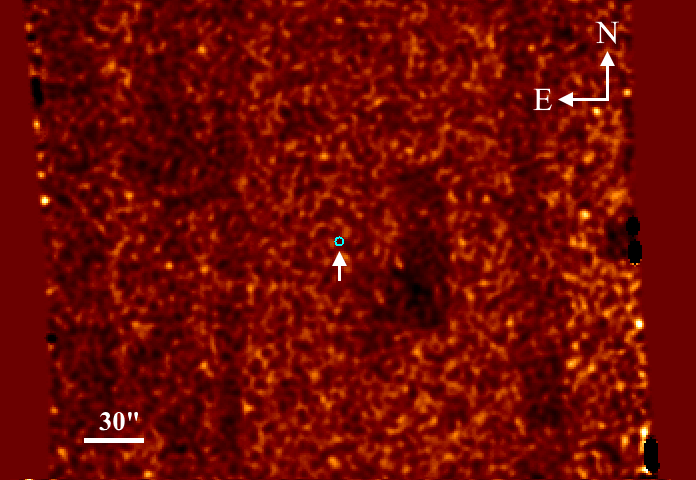}
\caption{A close-up view of the ZTF non-detection image of 2I in November 2018. The image is smoothed by a moving Gaussian function with a width of 7~pix at $1.5\sigma$ and is plotted in hyperbolic sine scale. The circle in cyan marks the nominal of the comet and the size of the $3\sigma$ uncertainty.}
\label{fig:nov2018}
\end{figure*}

A point worth addressing is the non-detection in November 2018. The $3\sigma$ uncertainty of the position is $\sim4''$ (4 pixels) along the major axis, and the motion rate difference between different orbit solutions is less than 1/4th of a pixel over the entire time span, but no comet is detected in the stack (Figure~\ref{fig:nov2018}). Subsequent light-curve analysis (see \S~\ref{sec:disc:driver}) shows that 2I should be $\sim1$~mag above than the $3\sigma$ limit of the stack.

\subsection{Pan-STARRS}

\begin{figure*}
\centering \includegraphics[width=0.8\textwidth]{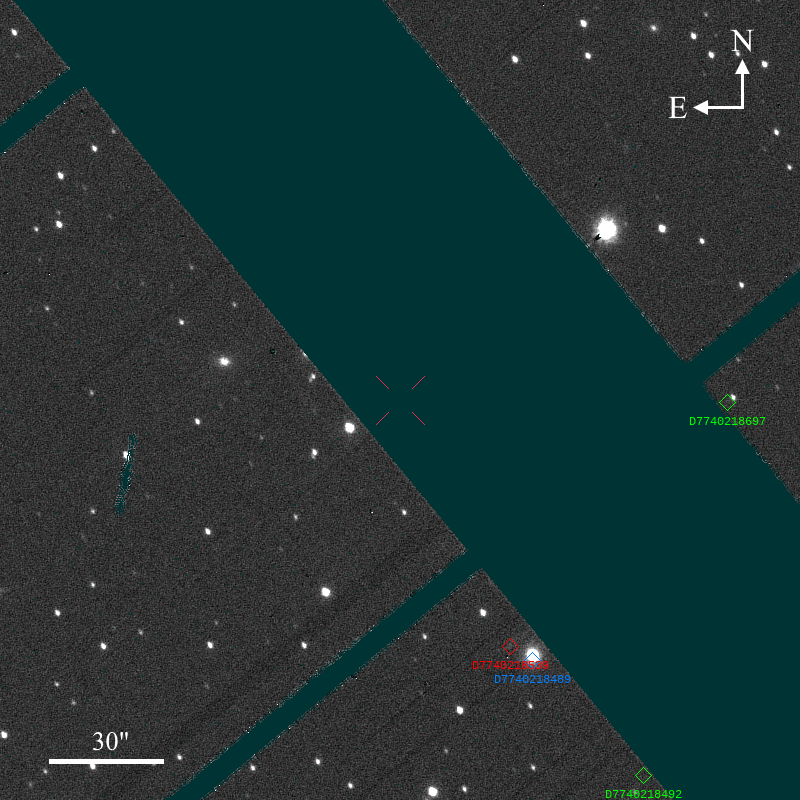}
\caption{Precovery image of 2I taken by PS1 on 2019 January 17. The comet's nominal location, marked by a red crosshair, is in the chip gap.}
\label{fig:ps1}
\end{figure*}

The Pan-STARRS survey \citep{Chambers2016} is a wide-field asteroid survey comprised of two identical 1.8 m telescopes, the Pan-STARRS1 (PS1) and Pan-STARRS2 (PS2). The survey has an image archive extending back to 2010. Using the ZTF precovery data along with the available astrometry from the Minor Planet Center up to 2019 October 1, we generate an ephemeris covering the period from 2018 January 1 (when 2I was well below any practical ground based telescope sensitivity) until the discovery date of 2019 August 30. Four 45 second $i$-band images taken by the PS1 system on 2019 January 17 were identified, as summarized in Table \ref{tbl:prediscovery}. The normal processing for Pan-STARRS applies masking to remove areas not optimal for photometry because of non-uniform charge transfer efficiency, and for this study, the exposures were reprocessed without this mask. They were then visually inspected carefully over a large region centered at the expected ephemeris. The predicted ephemeris and uncertainty, however, place 2I firmly into a $70''$-wide chip gap (Figure~\ref{fig:ps1}).

While no detections were found in the Pan-STARRS pre-discovery images, the fact that the expected location of 2I is contained within a chip gap favors the ZTF precovery positions being correct, as even a weakly active comet should have been visible otherwise given the $G \sim 23$ limiting magnitude and near arcsecond seeing. The chip gap only occupies a small fraction ($\sim10\%$ based on the astrometry up to 2019 October 1) of the uncertainty ellipse, therefore there was a much larger chance of being proven wrong.

However, we note that PS1 was not operational between 2018 August 23 and December 12 due to a dome shutter failure, and between 2019 February 10 and March 27 due to loss of power after a winter storm on Haleakala where the telescope is installed.


\subsection{Catalina Sky Survey}

The Catalina Sky Survey \citep[CSS; e.g.][]{Christensen2018} operates three telescopes dedicated to the discovery and follow-up of near-Earth objects: the 0.7 m Catalina Schmidt at Mt. Bigelow, a 1.0 m telescope and a 1.5 m telescope at Mt. Lemmon, Arizona.

Following a similar strategy of the Pan-STARRS data search, we searched the archival data of all three telescopes dating back to 2018 January 1. We identified various sets of images covering the predicted ephemeris of 2I: multiple sets of images taken by the 0.7 m Catalina Schmidt, but all dated no later than 2018 December 15, and a set of four images taken by the 1.5 m telescope on 2019 March 1.

\begin{figure}
\center\includegraphics[width=0.3\textwidth]{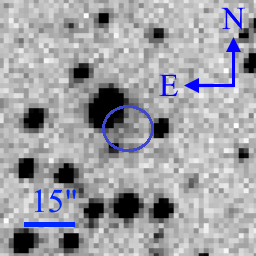}
\caption{Precovery image of 2I taken by the CSS 1.5 m telescope on 2019 March 1. The slight enhancement possibly corresponds to the comet is marked by a blue circle. The color is inverted.}
\label{fig:g96}
\end{figure}

The comet was too faint ($\gtrsim1.5$~mag beyond the limit of the image) for the December 2018 and earlier images, while the March 2019 images reached a limiting magnitude that could be compatible with a faint detection of 2I. However, the position corresponding to the ZTF detections fell on a region heavily contaminated with background field sources. In three of the four frames, the object would have overlapped the PSF of a field star. The remaining frame was also marginally affected by an even brighter nearby star, but it might show a slight enhancement that is within 1.5--2 pixels from the predicted position of 2I (Figure~\ref{fig:g96}). A deeper stack of historical images obtained by Catalina with the same telescope reveals no background source at that position, down to a limiting magnitude much fainter than the individual frame. Unfortunately, the enhancement was extremely faint, and might be compatible with a noise feature enhanced by the tails of the bright star's point-source-function (PSF), making it difficult to draw a solid conclusion.

\section{Photometric Analysis}

To measure the flux from 2I for each epoch specified in the 2nd column of Table~\ref{tbl:prediscovery}, we scale each frame in the time range listed in the 1st column of the table with respect to a ``reference'' frame in this range using the following formula for frame combination:

\begin{equation}
\label{eq:scale}
    S = 10^{-0.4{mag_\mathrm{ZP, corr}}} \left( \frac{\varDelta}{\varDelta_0} \right)^2 \left( \frac{r_\mathrm{H}}{r_\mathrm{H, 0}} \right)^4 
\end{equation}

\noindent where $S$ is the scale coefficient; $\varDelta$ and $\varDelta_0$ is the geocentric distance of the comet in the given frame and the reference frame, respectively; $r_\mathrm{H}$ and $r_\mathrm{H, 0}$ is the heliocentric distance of the comet in the given frame and the reference frame, respectively\footnote{The exponent terms of $\varDelta$ and $r_\mathrm{H}$ come from the classic comet brightness formula, $m_1=M_1 + 5 \log{\varDelta} + 10 \log{r_\mathrm{H}}$, where $m_1$ and $M_1$ are the apparent and absolute total magnitude of the comet, and the flux of the comet is proportional to $\varDelta^{-2}$ and $r_\mathrm{H}^{-4}$, respectively. If we leave out these two terms, the non-detection photometry will differ by 0.5 mag compared to the values listed in Table~\ref{tbl:prediscovery}. For other pre-discovery data points, the differences are within 0.05~mag.} and the corrected magnitude zero-point $mag_\mathrm{ZP, corr}$ is defined by 

\begin{equation}
    mag_\mathrm{ZP, corr} = mag_\mathrm{ZP} + C_\mathrm{img} \times C_\mathrm{comet}
\end{equation}

\noindent where $mag_\mathrm{ZP}$ is the magnitude zero-point and $C_\mathrm{img}$ is the color coefficient derived by the ZTF Science Data System and calibrated to the PS1 photometric system \citep{Masci2019}, and $C_\mathrm{comet}$ is the color of the comet. We use $C_{\mathrm{comet:} g-r}=g_\mathrm{PS1}-r_\mathrm{PS1}=0.54$ as measured by \citet{Guzik2019}, with the colors converted using the relations derived in \citet{Tonry2012}. We use a fixed photometric aperture with a radius of 5 pixels, or 21000--29000~km at the comet in the interval of November 2018 and May 2019. This aperture is sufficient to include all flux from the comet in the pre-discovery data. The result is tabulated in Table~\ref{tbl:prediscovery}.

We then fit the light-curve using the classic comet light-curve equation:

\begin{equation}
\label{eq:m1}
    m_1 = M_1 + 5 \log{\varDelta} + K_1 \log{r_\mathrm{H}} + \Phi(\alpha)
\end{equation}

\noindent where $m_1$ and $M_1$ are the apparent and absolute total magnitude of the comet, respectively, $K_1$ is the logarithmic heliocentric distance slope, and $\Phi(\alpha)$ is the phase function of the comet with respect to the phase angle $\alpha$. We note that this process should not be confused with the image scaling process with Equation~\ref{eq:scale} as described above, as the purpose of the image scaling process was to scale a subset of data to a reference epoch (defined in the 2nd column of Table~\ref{tbl:prediscovery}), while the photometry of each reference epoch is then used for the light-curve fitting described here.

\begin{figure}
\center\includegraphics[width=0.5\textwidth]{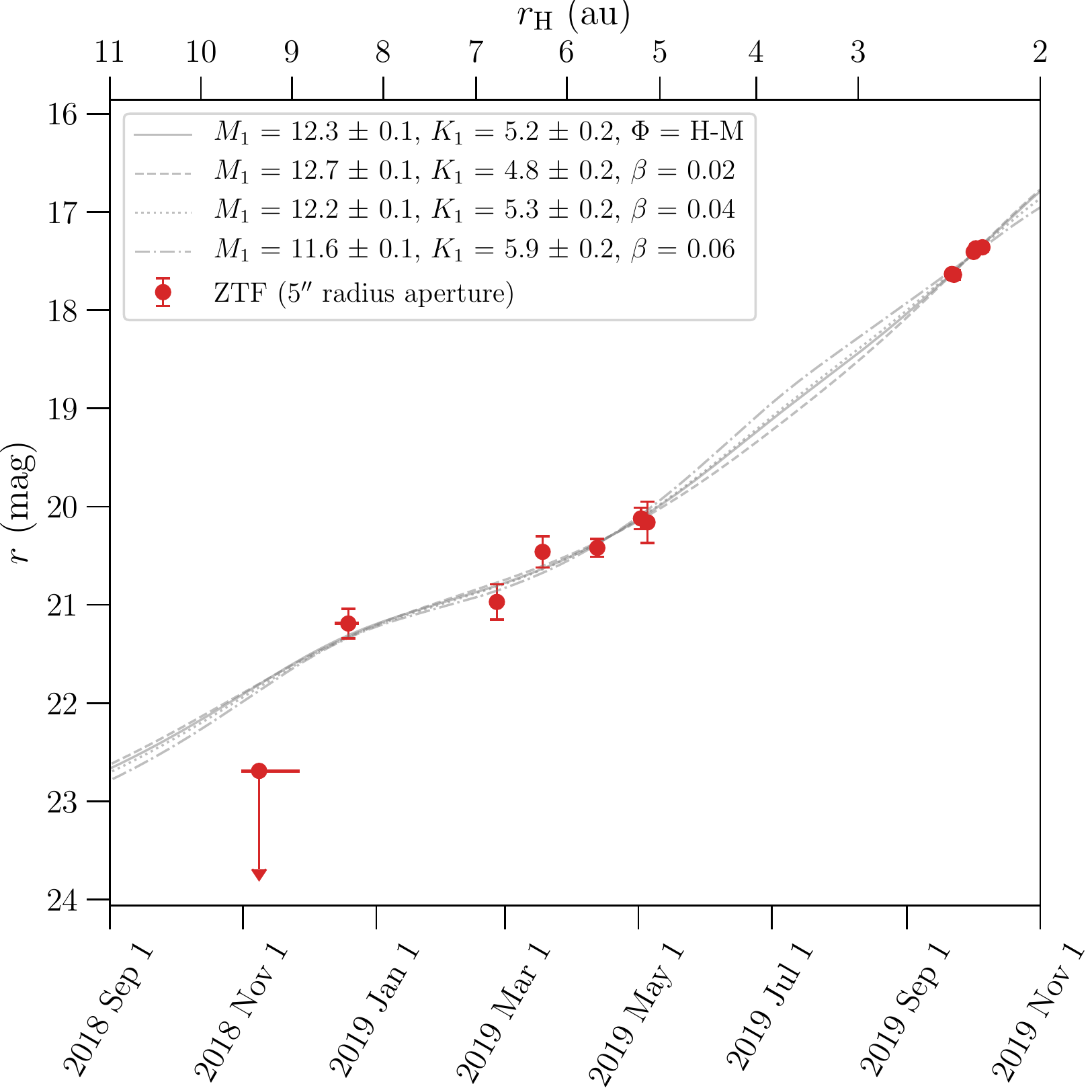}
\caption{Best-fit light-curve of 2I using the pre-discovery photometry and recent ZTF photometry (\href{https://www.minorplanetcenter.net/mpec/K19/K19V34.html}{MPEC 2019-V34} and \href{https://www.minorplanetcenter.net/mpec/K19/K19W50.html}{MPEC 2019-W50}). Horizontal bars indicate the time bin sizes. The light-curve functions being fitted are the Marcus model on Halley's Comet (H-M), as well as linear phase functions assuming a phase coefficient of 0.02, 0.04 and 0.06 mag~deg$^{-1}$. The bumps in December 2018 and July 2019 are caused by brightness enhancements at small phase angles during the opposition and conjunction of 2I.}
\label{fig:lc}
\end{figure}

We test four phase functions: the \citet{Marcus2007} model on Halley's Comet, as well as linear phase functions $\Phi(\alpha)=\alpha \beta$ where $\beta=0.02$, 0.04, and 0.06~mag~deg$^{-1}$ is the phase coefficient of the comet \citep{Lamy2004}. All four models yield comparable results, with $M_1$ varies from $11.6\pm0.1$ to $12.7\pm0.1$, and $K_1$ from $4.8\pm0.2$ to $5.8\pm0.2$, as shown in Figure~\ref{fig:lc}.

\section{Discussion}

\subsection{Driver of the Activity}
\label{sec:disc:driver}

The best-fit light-curve as shown in Figure~\ref{fig:lc} revealed a shallow slope, $K_1=5.3\pm0.2$, taking a linear phase coefficient $\beta=0.04$~mag~deg$^{-1}$. A shallow slope means that the activity stays largely constant as the comet approaches the Sun. \citet{Whipple1978} shows that shallow brightening is common on the ``dynamically new'' solar system comets (with orbital periods $P\gg10^4$~yr) which have $K_1 \in (5, 8)$), but uncommon on short-period comets and other long-period comets, which have $K_1\gtrsim10$, though his dataset is dominated by small heliocentric distances with $r_\mathrm{H}<5$~au. In this respect, 2I is analogous to dynamically new comets in the solar system.

The $K_1=5.3$ slope seemingly deviates from the data beyond $\sim8$~au, preceded by what appears like a steep brightening phase, with $K_1\approx29$, though additional pre-discovery observations are needed to verify it. If such steep phase is real, it might indicate the onset of sublimation of cometary volatiles. The most compatible major volatile would be CO$_2$, which has an onset ``knee'' distance $r_\mathrm{H}=13$~au. Other cometary species, such as CH$_3$CN, HCN and CH$_3$OH, have similar turn-on distances \citep{Meech2004}, but they have low abundances in solar system comets \citep{Cochran2015}. CO, another cometary volatile commonly found on dynamically new comets in solar system that has an onset distance $r_\mathrm{H}=120$~au, is not as compatible. However, it has been suggested for the case of `Oumuamua that the volatiles may be buried beneath the surface and are only activated when the comet is much closer to the Sun, due to the time lag for the heat wave to penetrate to the depth of the ice \citep{Fitzsimmons2018, Seligman2018}. Therefore, it is too early to exclude CO as the main driver of 2I's activity.

An alternative explanation of the activity is the exothermic crystallization of amorphous water ice, a mechanism that may be responsible for the activity of distant comets \citep{Prialnik1990}. Amorphous ice forms below an environmental temperature of $\sim130$~K and is capable of trapping gas as they form, a phenomenon that has been observed in laboratory experiments \citep{BarNun1987, Jenniskens1994}, though the presence of amorphous ice is yet to be directly observed on comets. Depending on the illumination of the cometary nucleus, crystallization of amorphous ice on the surface can start around $6$--$12$~au \citep{Jewitt2017}, consistent with the observed turn-on distance of 2I.

To gain a deeper insight into the light-curve, we tested a sublimation model \citep{Meech2004} that computes the amount of gas sublimating from an icy surface exposed to solar heating to explore the activity. The total brightness within a fixed aperture combines radiation scattered from both the nucleus and the dust dragged from the nucleus in the escaping gas flow, assuming a dust to gas mass ratio of 1. 
We used a nucleus radius of 0.5 km \citep{Jewitt2019b}, assumed an albedo of 0.04 for the nucleus and a linear phase function of 0.04 mag~deg$^{-1}$ for the nucleus and 0.02 mag~deg$^{-1}$ for the coma typical of other comets \citep{Meech1987,Krasnopolsky1987}, a nucleus density of 400 kg~m$^{-3}$ similar to that seen for comets 9P/Tempel 1, 103P/Hartley 2 \citep{thomas2009} and 67P/Churyumov-Gerasimenko \citep{patzold2016}, a grain density of 800 kg~m$^{-3}$ \citep{Fulle2016}, and large (10--100~$\mu$m) grains \citep{Fitzsimmons2019}.
Unsurprisingly, our model confirmed that the activity of 2I must be driven by a species more volatile than water, since otherwise it would have been well below the detection limit of any of the surveys at $\sim5$~au. We also found that the differences between the shape of the sublimation curves for CO and CO$_2$ near 8~au is minimal, so it is impossible to distinguish between these volatiles without further pre-discovery observations. 

\subsection{Size of the Nucleus}
\label{sec:disc:nucleus}

The non-detection in November 2018 data can be used to constrain the size of the nucleus. The effective cross-section area for scattering can be calculated using

\begin{equation}
\label{eq:ce}
    C_\mathrm{e} = \frac{\pi (1~\mathrm{au})^2}{p} 10^{-0.4 (M_r - m_{\odot, r})}
\end{equation}

\noindent where $p=0.04$ is the assumed optical geometric albedo of the nucleus \citep{Lamy2004}, $m_{\odot, r}=-26.9$ is the apparent $r$-band magnitude of the Sun \citep{Willmer2018}, and the absolute $r$-band magnitude (technically, the upper limit) $M_r$ is defined as

\begin{equation}
    M_r = m_r - 5\log{r_\mathrm{H} \varDelta} - \Phi{(\alpha)}
\end{equation}

\noindent with the variables following the same definitions for Equation~\ref{eq:m1}. By inserting all the numbers, we have $C_\mathrm{e}<140~\mathrm{km^2}$ for the non-detection in November 2018, at 8.5~au. This upper bound indicates that the radius of the nucleus is no larger than $\sim7$~km.

\subsection{Active Area on the Nucleus}

The size of the active area on the nucleus can be estimated with the knowledge of the mass loss rate of the comet and the mass flux of the activity-driving volatiles. The mass loss rate of 2I can be estimated using the cross-section area of the dust and the speed of the dust flow. By using Equation~\ref{eq:ce}, we derive the cross-section area of coma to be $\sim280$ to $360~\mathrm{km^2}$, from March to October 2019, in which $r_\mathrm{H}$ decreases from 6.4 to 2.6~au. The mass loss rate $\dot{M}$ can be calculated by

\begin{equation}
    \dot{M} = \frac{4 \rho \bar{a} C_\mathrm{e}}{3 \tau}
\end{equation}

\noindent where $\rho=800~\mathrm{kg~m^{-3}}$ is the bulk density of the dust \citep{Fulle2016}, which is admittedly not yet constrained for interstellar comets, but we do not have reason to believe it is much different from solar system comets, and therefore have assumed it to be analogous to the latter, $\bar{a}=10^{-4}~\mathrm{m}$, the characteristic size of the dust, is similarly assumed based on the observation of dynamically new solar system comets \citep[e.g.,][]{Ye2014,Jewitt2019}, and $\tau$ is the timescale that a dust particle moves out of the aperture, which can be estimated by $\tau=l/v\approx 10^5~\mathrm{s}$, where $l$ is the linear length of the aperture at the comet, and $v\sim1~\mathrm{m~s^{-1}}$ is the ejection speed of the dust, taking the dust speed constrained by \citet{Guzik2019} and assuming a classic $\propto a^{-0.5}$ dependence. By inserting all the numbers, we obtain $\dot{M}\approx10~\mathrm{kg~s^{-1}}$, with the uncertainty around an order of magnitude mainly owing to the parameter $\bar{a}$ \citep[which can vary by an order of magnitude among solar system comets, cf.][]{Fulle2004}.

We then solve the energy balance equation for CO and CO$_2$ ice, which are likely to be responsible for 2I's activity following the discussion in \S~\ref{sec:disc:driver}, at the sub-solar point on the nucleus:

\begin{equation}
    \frac{(1-A_B) L_\odot}{4 \pi r_\mathrm{H}^2} = \epsilon \sigma T^4 + Z(T) L(T)
\end{equation}

\noindent where $A_B=0.01$ is the Bond albedo of the nucleus measured for 9P/Tempel 1 and 103P/Hartley 2 \citep{Li2013, Li2013b}, $L_\odot$ is the luminosity of the Sun, $\epsilon=0.9$ is the infrared emissivity of the nucleus, $\sigma$ is the Boltzmann constant, $L(T)$ is the latent heat of the sublimation of the ice, and $Z(T)$ is the mass flux. We solve $L(T)$ using the model by \citet{Cowan1979}\footnote{\url{https://pdssbn.astro.umd.edu/SBNcgi/newiso.cgi}}, and obtain $Z(T)\approx3\times10^{-6}$---$2\times10^{-5}~\mathrm{kg~m^{-2}~s^{-1}}$ from 6.4~au 2.6~au for CO, and $1\times10^{-6}$---$9\times10^{-6}~\mathrm{kg~m^{-2}~s^{-1}}$ for the same $r_\mathrm{H}$ span for CO$_2$. The active area required to support the mass loss rate would then be

\begin{equation}
    A = \frac{\dot{M}}{(M_\mathrm{d}/M_\mathrm{g}) Z(T)}
\end{equation}

\noindent where $M_\mathrm{d}/M_\mathrm{g}$ is the dust-to-gas mass ratio, which is again unknown for interstellar comets. If we take $M_\mathrm{d}/M_\mathrm{g}=1$ based on the measurement of long-period comet C/1995 O1 (Hale-Bopp) at similar heliocentric distance \citep{Weiler2003}, we have $A=0.5$---$10~\mathrm{km^2}$ over the time span between December 2018 and September 2019, in line with the number derived by \citet{McKay2019} based on their observation in October 2019 ($1.7~\mathrm{km^2}$). Taking the upper limit of the size of the nucleus derived in \S~\ref{sec:disc:nucleus}, the active fraction of 2I is $>0.1\%$ of the nucleus. This is in line with the known solar system comets, which have fractional active area from a few 0.1\% to $>100\%$ \citep{Tancredi2006}, though we note that most of these measured comets are short-period comets with H$_2$O as the driving species. However, we caution that the uncertainty in $A$ is about 1--2 orders of magnitude when we consider the uncertainties in $\dot{M}$ and $M_\mathrm{d}/M_\mathrm{g}$, therefore the derived $A$ and active fraction is highly uncertain.

\subsection{Non-gravitational Accelerations and Implications}
\label{sec:disc:nongrav}

The inclusion of the precovery data in the orbit estimation process provides more stringent constraints on the trajectory of 2I, but also introduces challenges to correctly model the dynamics. A gravity-only model of the orbit struggles to fit the data of March 2019 and earlier. In particular, the December 2018 position is rejected as an outlier \citep[using the outlier rejection algorithm by][]{carpino03}. At this stage it is not entirely clear whether this behavior is caused by systematic errors in the bulk of the astrometric data, which were taken at low solar elongation and therefore at high airmass, or by non-gravitational accelerations. We note that non-gravitational accelerations were detected in the motion of `Oumuamua, despite the lack of visible outgassing \citep{Micheli2018}.

Table~\ref{tbl:solutions} reports JPL solution 37, which fits all the precovery observations and uses non-gravitational forces assuming CO as the primary driver (\S~\ref{sec:predisc:ztf}), more consistent than H$_2$O with the photometric data. The non-gravitational model for CO$_2$ is not available at this point; but since both CO and CO$_2$ are more volatile than water and that 2I was in the regime of both volatiles when it was observed, we believe that the two models should behave similarly over the fit span.

We also tested the rotation jet model \citep{chesley05}, which computes the non-gravitational perturbations from a discrete number of jets, whose acceleration is averaged over a nucleus rotation. For the driver of the activity we again used CO. We considered two jets, a nearly polar one at 10$^\circ$ of colatitude and a mid-latitude one on the southern hemisphere at 135$^\circ$ of colatitude. Then, we scanned a raster for the spin pole's Right Ascension (RA) and Declination (Dec), estimating the strengths of the two jets from the fit to the astrometry.

\begin{figure}
\center\includegraphics[width=0.5\textwidth]{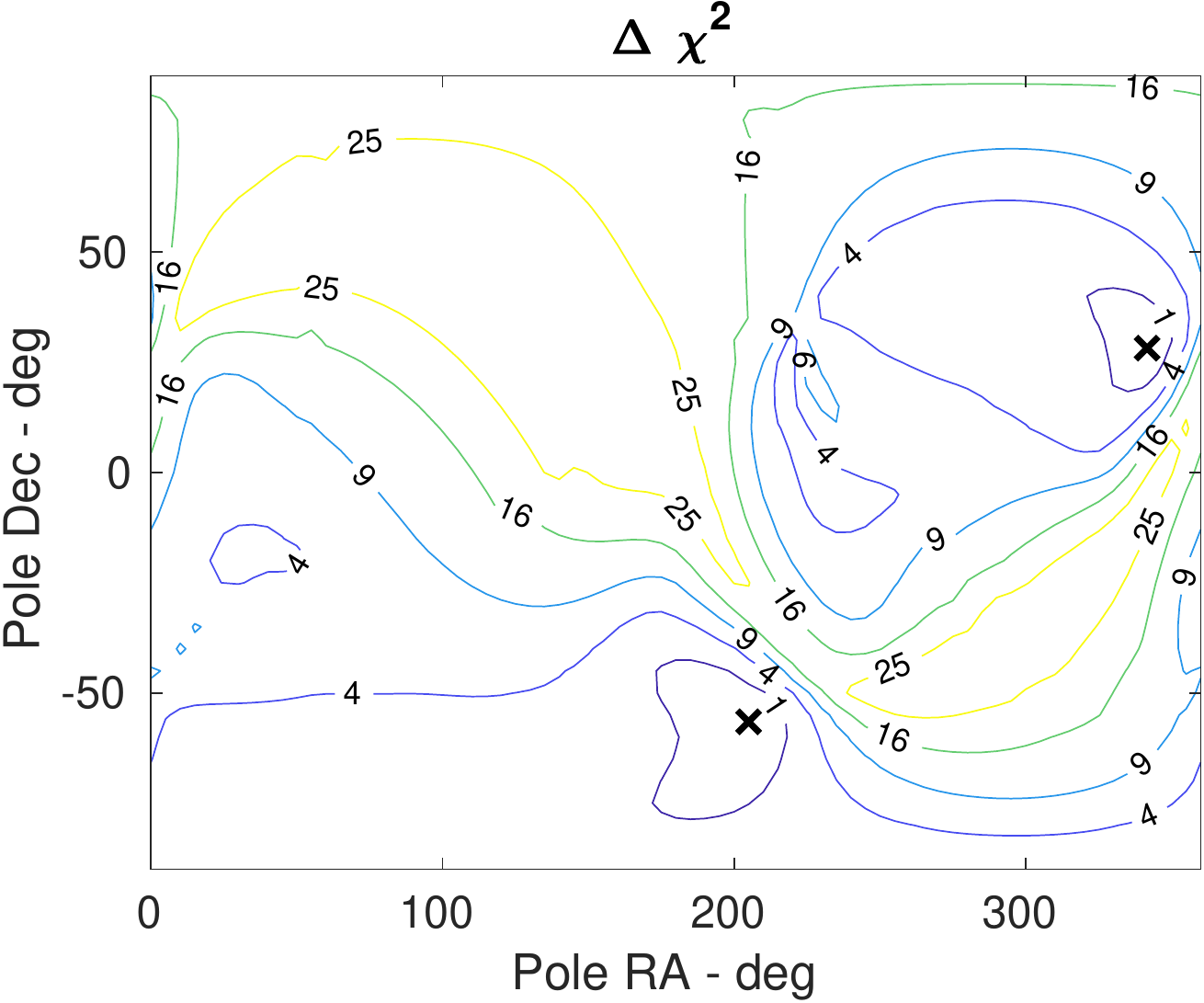}
\caption{$\Delta\chi^2$ of the astrometric fit for the jet model as a function of the pole orientation. The two minima are marked with a black cross.}
\label{fig:pole-rjm}
\end{figure}

As shown in Fig.~\ref{fig:pole-rjm}, we find two minima for the pole's RA and Dec: ($340^\circ$, $+30^\circ$) and ($205^\circ$, $-55^\circ$). Because of the larger number of parameters, the jet model provides a better fit to the data than the \citet{Marsden1973} model. However, its reliability will need to be validated by the capability of making accurate predictions. Past experience has shown that the jet model can provide more accurate comet trajectory estimates \citep{farnocchia16} and therefore the jet model solutions are worth consideration, especially as the observed arc extends into late 2019 and 2020.

\section{Conclusions}

The pre-discovery observations of newly-discovered interstellar comet 2I/Borisov revealed a comet that is observationally quite comparable to the long-period dynamically new comets in our own solar system. We found that 2I was active at 5--7~au, indicative of the presence of accessible ices more volatile than H$_2$O, such as CO and CO$_2$. A subsequent comprehensive follow-up campaign, presented by \citet{Bolin2019}, reinforces this conclusion. We identified a possible steep brightening at 8--9~au might indicate an onset of activity at this distance, which suggests crystallization of amorphous ice as an alternative mechanism for the activity, but more pre-discovery data, preferably from larger, multi-meter-sized telescopes, is needed to verify this behavior. We also found the nucleus to be no more than 7~km in radius, and that $\gtrsim0.1\%$ of the surface is currently active, both are quite typical when compared to dynamically new solar system comets occasionally discovered and observed by surveys, though the derived size of active area is highly uncertain, mainly due to the uncertainties in nucleus size and dust size distribution. The pre-discovery observations also provides stronger constraints on the inbound trajectory and non-gravitational forces of 2I. We found that a CO model provides results that are more consistent with the observations compared to H$_2$O model.

It will be interesting to see if 2I continues to fit into the profile of dynamically new comets. For solar system comets, it is known that dynamically new comets are $10\times$ more likely to disintegrate than short-period comets, presumably due to their pristine state and weaker structural strength \citep{Weissman2004}. We note that independent analysis by \citet{Jewitt2019b} also suggested that 2I may be prone to disintegration based on its small nucleus size (sub-km-sized). Comets can disintegrate at large heliocentric distances, but most disintegrations seem to happen within $\sim2$~au \citep{Boehnhardt2004}, a distance that 2I will reach at its perihelion in December 2019. Survived dynamically new comets also tend to fade more rapidly after perihelion \citep{Whipple1978}. Continued observations of 2I will enable further comparison to dynamically new comets in our solar system, and provide timely warning for any disintegration (or, as a less dramatic form, outburst) that may happen.




\acknowledgments

The authors thank George Helou, Man-To Hui, David Jewitt, Matthew Knight, Zhong-Yi Lin, Ralph Roncoli, Qicheng Zhang, and an anonymous referee for helpful discussions and comments. MSPK acknowledges support from NASA grant NNX17AK15G. DF conducted this research at the Jet Propulsion Laboratory, California Institute of Technology, under a contract with NASA.

Based on observations obtained with the Samuel Oschin 48-inch Telescope at the Palomar Observatory as part of the Zwicky Transient Facility project. ZTF is supported by the National Science Foundation under Grant No. AST-1440341 and a collaboration including Caltech, IPAC, the Weizmann Institute for Science, the Oskar Klein Center at Stockholm University, the University of Maryland, the University of Washington, Deutsches Elektronen-Synchrotron and Humboldt University, Los Alamos National Laboratories, the TANGO Consortium of Taiwan, the University of Wisconsin at Milwaukee, and Lawrence Berkeley National Laboratories. Operations are conducted by COO, IPAC, and UW.

Pan-STARRS is supported by the National Aeronautics and Space Administration under Grant No. 80NSSC18K0971 issued through the SSO Near Earth Object Observations Program.

\software{ZChecker \citep{Kelley2019}}
\facilities{PO:1.2m, PS1, SO:1.5m}

\end{CJK*}
\bibliographystyle{aasjournal}
\bibliography{ms}



\end{document}